\documentstyle[prl,twocolumn,aps,floats,epsf,epsfig,amsmath,amssymb]{revtex}

\begin{document}

\twocolumn[
\hsize\textwidth\columnwidth\hsize\csname@twocolumnfalse\endcsname

\title{In-plane magnetodrag in dilute bilayer two-dimensional systems:
  a Fermi liquid theory} 
\author{S. Das Sarma and E. H. Hwang}
\address{Condensed Matter Theory Center, Department of Physics, University of
Maryland, College Park, Maryland 20742-4111}
\date{\today}
\maketitle

\begin{abstract}
Motivated by recent experimental results reporting anomalous drag
resistance behavior in dilute bilayer two-dimensional (2D) hole
systems in the presence of a magnetic field parallel to the 2D plane,
we have carried out a many-body Fermi liquid theory calculation of
bilayer magnetodrag comparing it to the corresponding single layer
magnetoresistance. In qualitative agreement with experiment we find
relatively similar behavior in our calculated magnetodrag and
magnetoresistance arising from the physical effects of screening being
similarly modified ("suppressed") by carrier spin polarization (at
"low" field) and the conductivity effective mass being similarly
modified ("enhanced") by strong magneto-orbital correction (at "high"
fields) in both cases. We critically discuss agreement and
disagreement between our theory and the experimental results,
concluding that the magnetodrag data are qualitatively consistent with
the Fermi liquid theory. 

\end{abstract}

\vspace{0.5cm}
]

Much attention has recently focused on low-density 2D systems in
semiconductor structures \cite{lilly,noh,zhu,yoon,pill1,kellogg,pill2}
where carrier transport properties may be
strongly affected by interaction effects. In particular, low
temperature transport \cite{lilly,noh}, magnetotransport
\cite{zhu,yoon}, drag \cite{pill1,kellogg}, and magnetodrag \cite{pill2}
properties have recently been studied in low-density electron
\cite{lilly,zhu,kellogg} and hole\cite{noh,yoon,pill1,pill2},
single-layer \cite{lilly,noh,zhu,yoon} and bilayer
\cite{pill1,kellogg,pill2} systems, providing a great deal of detailed
quantitative information on the temperature, density, and magnetic
field dependent 2D resistivity $\rho(T,n,B)$ and 2D drag-resistivity
$\rho_D(T,n,B)$ behavior. (The externally applied magnetic field being
discussed throughout this work and in the relevant experimental
references \cite{lilly,noh,zhu,yoon,pill1,kellogg,pill2} is an
`in-plane' magnetic field $B$ applied parallel to the 
2D layer.) A very recent experimental work \cite{pill2} by
Pillarisetty {\it 
  et. al.} reports some striking qualitative resemblance between the
2D bilayer drag $\rho_D$ and the corresponding single-layer
resistivity $\rho$ as a function of the applied parallel field $B$ in
a low density low disorder 2D GaAs hole system. Since the physical
mechanisms underlying $\rho$ and $\rho_D$ are generally thought to be
qualitatively different at low temperatures, the experimental
observations of ref. 7 take on important qualitative significance. In
particular, $\rho$ in high-mobility 2D systems at low temperatures is
entirely due to scattering by random charged impurity centers whereas
$\rho_D$ at low temperatures arises entirely from inter-layer
electron-electron scattering. (Electron-phonon scattering makes
negligible contributions to both $\rho$ and $\rho_D$ at low
temperature \cite{lilly,noh}.) Since electron-electron scattering does
not directly 
contribute to $\rho$ in translationally invariant 2D semiconductor
systems, the reported \cite{pill2} qualitative similarity between the
observed 
$\rho$ and $\rho_D$ behaviors in its magnetic field dependence
presents a significant theoretical challenge. Since it is manifestly
obvious that electron-impurity scattering can at best play an
unimportant and indirect secondary role \cite{drag} in determining the
interlayer drag resistance, the experimental observation of ref. 7
raises very serious fundamental questions regarding our understanding
of the nature of the ground state of a low-density 2D carrier
system. We note that electron-electron interaction induced umklapp
scattering, which could contribute 
to the single-layer resistivity (since umklapp processes do not
conserve momentum), is completely irrelevant in 2D semiconductor
structures where all the electronic physics occurs essentially at the
zone-center $\Gamma$ point in the effective mass approximation sense
(and the real lattice structure does not play any role).

In view of the considerable fundamental significance of the issues
raised by the experimental observations, we present in this Letter a
careful theoretical calculation of both $\rho(B)$ and $\rho_D(B)$ in a
low-density 2D carrier system within the canonical many-body Fermi
liquid theory that has earlier been found to be successful in
providing a reasonable qualitative (and perhaps even
semi-quantitative) description of the approximate temperature and
density dependence of $\rho$ \cite{dassarma1} and $\rho_D$
\cite{hwang1} at low temperatures and
densities {\it in the absence of} any applied in-plane magnetic
field. We note that the zero-field temperature and carrier density
dependence of 2D resistivity $\rho(T,n)$ and 2D drag resistivity
$\rho_D(T,n)$ {\it in the absence of any external magnetic field} are
certainly very different as one would expect on the basis of $\rho$ and
$\rho_D$ being determined by different scattering processes: $\rho$ by
screened charged impurity scattering and $\rho_D$ by interlayer
electron-electron scattering. For example, $\rho$ shows
\cite{lilly,noh,zhu,yoon} an approximate
linear increase with $T$ at low temperatures as is expected
\cite{dassarma1} for screened
Coulomb impurity scattering and $\rho_D$ shows
\cite{pill1,drag,hwang1} an approximate quadratic
increase with $T$ at low temperatures as is expected for
electron-electron scattering. (Similarly the carrier density dependence
of $\rho$ and $\rho_D$ are also very different at $B=0$.) The question
therefore naturally arises why the in-plane magnetic field dependences
of $\rho(B)$ and $\rho_D(B)$ reported in ref. [7] show qualitative
similarities. 

We theoretically argue, showing concrete calculated result within the
many body Fermi liquid theory, that the qualitative magnetic field
dependence of $\rho(B)$ and $\rho_D(B)$ should indeed be similar since
the scattering processes controlling the two properties (electron
charged impurity scattering for $\rho$ and electron-electron
scattering for $\rho_D$) are both screened by the carriers themselves
and the dominant behavior in both cases arises primarily from the
magnetic field dependence of electronic screening [11] (through the
spin polarization process) and (somewhat to a lesser degree) from the
magneto-orbital effect [12] (through the modifications of the 2D
conductivity effective mass and the confining quasi-2D
wave function). The reported qualitative similarity between magnetodrag
and magnetoresistance thus arises from drag and resistance being
dominated by screened carrier-carrier scattering and screened
carrier-impurity scattering respectively. The fact that 
{\it long-ranged charged}
impurity potential is the dominant source of resistive scattering in 2D
semiconductor structures (and this {\it long-ranged charged} 
impurity scattering must
necessarily be screened by the carriers) is therefore the key reason
for the broad qualitative similarity between $\rho_D(B)$ and $\rho(B)$
reported in [7].

We start by writing down the zero-field theoretical formulae for
$\rho$ [9] and $\rho_D$ [10,8] in the many-body Fermi liquid
RPA-Boltzmann theory approximation widely used in the literature.
The resistivity is given by $\rho^{-1} = ne^2 \langle \tau \rangle/m$,
where $n$, $m$ are the 2D 
carrier density and the conductivity effective mass respectively
whereas the transport relaxation time $\tau$ is given by
\begin{equation}
\frac{1}{\tau(\varepsilon_k)}=\frac{2\pi}{\hbar}\sum_{{\bf k}'}
n_i |u_{ei}({\bf k}-{\bf k}')|^2
(1-\cos\theta_{\bf {kk}'})\delta(\varepsilon_k-\varepsilon_{k'}) ,
\end{equation}
with $\langle\tau\rangle$ being a thermal average over the carrier
energy $\varepsilon$. Here $n_i$ is the density of charged impurity
centers in the 2D system (including the interface and the insulator),
and $u_{ei}(q)$ is the screened carrier-impurity scattering strength
give by $u(q)=v^c(q)/\epsilon(q)$, where $\epsilon(q)$ is the
single-layer 2D carrier dielectric function. (For details on the
derivation and implications of the formula for $\rho$, see ref. [9].)
The drag resistivity is given by 
\begin{equation}
\rho_D=\frac{\hbar^2}{2 \pi e^2n^2 k_BT}\int\frac{q^2 d^2q}{(2\pi)^2}
\int\frac{d\omega}{2\pi} \frac{F_1(q,\omega)F_2(q,\omega)}
{\sinh^2(\beta \omega/2)},
\end{equation}
where $F_{1,2}(q,\omega) = |u_{12}^{sc}(q,\omega)| \rm{Im}
\Pi_{11,22}(q,\omega)$, with $u_{12}^{sc} = v_{1122}^c
/\epsilon(q,\omega)$ is the dynamically screened interlayer Coulomb
interaction between layers 1 and 2, and $\Pi$ is the 2D
polarizability. (We consider the so-called balanced situation here with
the same carrier density $n$ in both layers.) Note that the dielectric
function ${\mathbf \epsilon}(q,\omega) = 1- v(q) \Pi(q,\omega)$ entering
Eq. (2) is the two component dielectric tensor for
the bilayer system [13]. (For details on the drag formula and its
implications, see refs. [8,10].)

It is important to emphasize that dielectric screening by the carriers
themselves is a key ingredient in determining both $\rho$ and $\rho_D$
although the static single layer (scalar) dielectric function
$\epsilon(q)$ determines $\rho$ through the screened charged impurity
scattering whereas the dynamical bilayer (tensor) dielectric function
$\epsilon(q,\omega)$ determines $\rho_D$ through the screened
interlayer Coulomb interaction. At low carrier densities used in
ref. [7], the difference between static and dynamic screening is not
of any qualitative significance since the effective plasma frequency
scale is rather low at low densities. Therefore both $\rho$ and
$\rho_D$ depend on the carrier dielectric function properties, which
is why they have qualitative similar magnetic field dependence as we
show and discuss below.

We have carried out a thoroughly nontrivial generalization of the above
theories for $\rho$ and $\rho_D$ to the finite in-plane magnetic field
situation $\rho(B)$, $\rho_D(B)$. Details will be provided elsewhere
[14], but here we mention the main physical effects of the applied
field for $\rho$ and $\rho_D$. The applied field has two completely
different physical effects through its coupling to carrier spin
(``magneto-spin'') [11] and orbital (``magneto-orbital'') dynamics
[12]. The magneto-spin effect arises from field-induced carrier spin
polarization due to the Zeeman coupling, and saturates at a density
dependent saturation field $B_s$ when the carrier system is fully spin
polarized (i.e. the magneto-spin effect exists only for $B\le
B_s$). The magneto-orbital effect [12] arises from the orbital
coupling of the in-plane magnetic field to the transverse dimension
due to the quasi-2D nature of the 2D layer and the magneto-orbital
effect is therefore monotonically increasing with increasing magnetic
field since this orbital coupling is important only when the magnetic
length $l=(c\hbar/eB)^{1/2}$ is smaller than the quasi-2D width of the
2D system. Thus one important qualitative difference between the
magneto-spin and the orbital effect is that the spin effect is
essentially a ``weak-field'' effect lasting only upto the saturation
field $B_s$ whereas the magneto-orbital effect increases monotonically
with increasing field.

The magneto-spin mechanism itself has two distinct effects: Suppression
of screening due to spin polarization [11] as the spin degeneracy
decreases from 2 (at $B=0$) to 1 (at $B \ge B_s$) and the increase of
the effective 2D Fermi surface as the value of the 2D Fermi wave
vector $k_F$ increases by a factor of $\sqrt{2}$ with $B$ increasing
from zero to $B_s$ due to the lifting of the spin
degeneracy. Similarly, the magneto-orbital mechanism also has two
distinct effects: The increase of the transport effective mass in the
direction perpendicular to the magnetic field direction and the
field-induced intersubband scattering among the quasi-2D subband ---
both of these are only operational at relatively high fields when
$l<a$ where $a$ is the average transverse width of the carrier
wave function. It is important to realize that three of these four
field induced effects (spin polarization induced screening
suppression, and both of the magneto-orbital effects) always produce
positive magnetoresistance whereas the Fermi surface effect (which is
significant only at high carrier densities where $2k_F \gg q_{TF}$,
$q_{TF}$ being the screening wave vector) always produces a negative
magnetoresistance. For the hole-doped  low-density samples of
ref. [7], the Fermi surface (i.e. $k_F \rightarrow \sqrt{2}k_F$ as
$B\rightarrow B_s$) effect is negligible since the system is in the
strong screening $q_{TF} \gg 2k_F$ limit.

\begin{figure}
\epsfysize=1.8in
\centerline{\epsffile{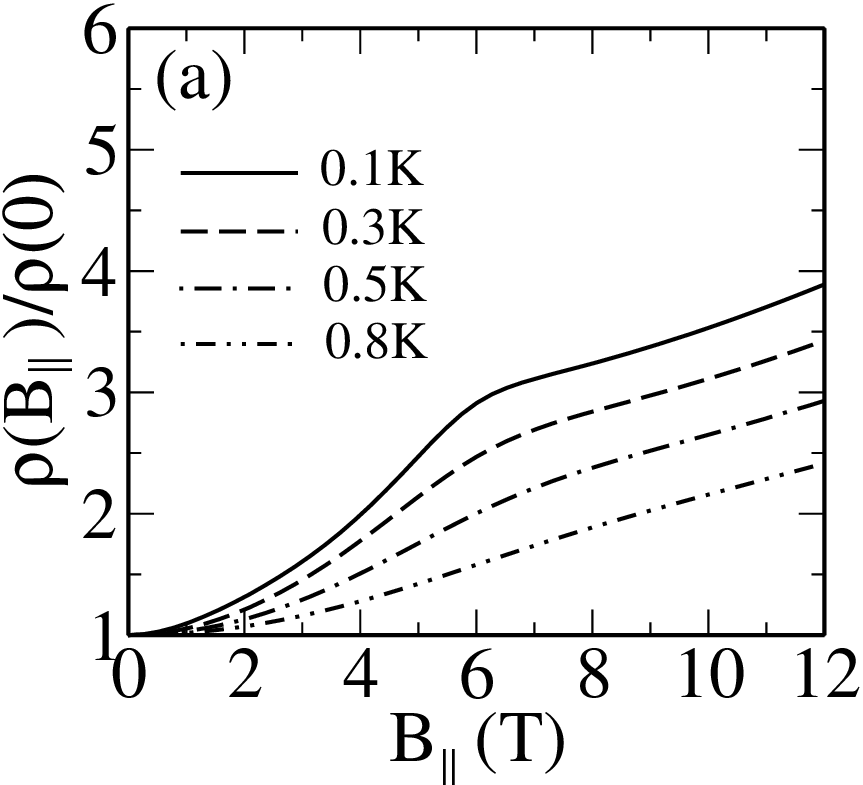}}
\centerline{\epsffile{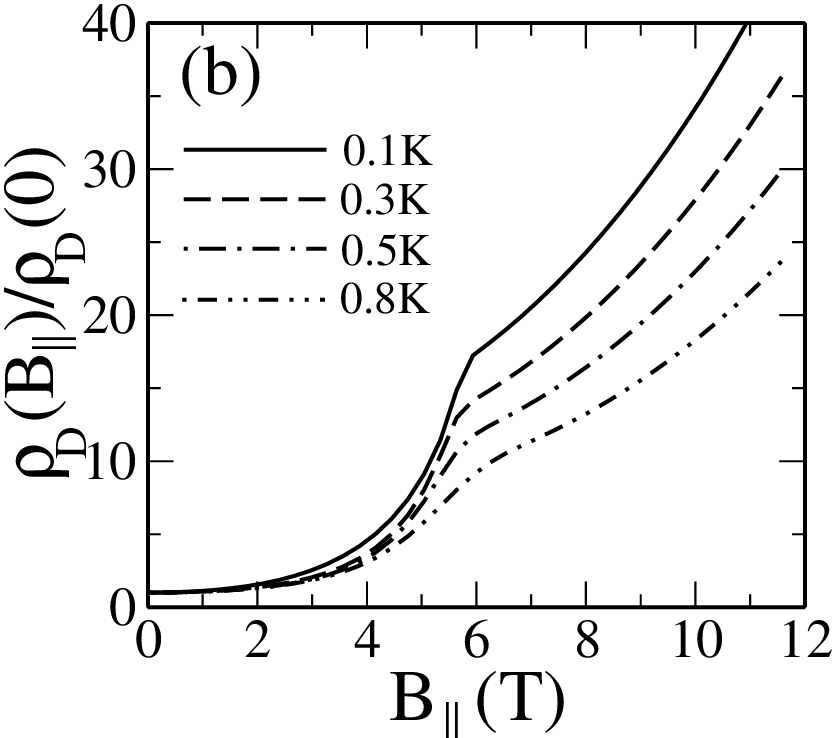}}
\caption{Calculated (a) magnetoresistance $\rho(B_{\|})$ and (b)
  magnetodrag $\rho_D(B_{\|})$ for hole density $p=2.15 \times 10^{10}
  cm^{-2}$ at various temperatures as a function of parallel field
  $B_{\|}$.  
\label{Fig1}
}
\end{figure}

The combination and the interplay of magneto-spin and magneto-orbital
effects are quite complex and sensitive to the parameter ($n,T,B$)
details, but a few general comments can still be made: (1) At low 
carrier densities [7] of interest to us, the static and dynamic
screening operational respectively in $\rho(B)$ and $\rho_D(B)$ behave
similarly, and therefore the spin-polarization induced screening effect
is qualitatively similar for $\rho(B)$ and $\rho_D(B)$; (2) since
field-induced magneto-spin effect operates only for $B \le B_s$, both
$\rho(B)$ and $\rho_D(B)$ manifest a cusp-type structure at $B=B_s$
where spins are completely polarized; (3) the maximum theoretically
allowed magnetoresistance $\rho(B)/\rho(0)$ and magnetodrag
$\rho_D(B)/\rho_D(o)$ corrections arising from the spin polarization
induced magneto-screening mechanism are factors of 4 and 16
respectively since screening itself could be suppressed at most by a
factor of 2 due to spin polarization effect (and Eqs. 1 and 2
respectively for $\rho$ and $\rho_D$ come with the second and the
fourth power of the spin degeneracy); (4) the main magneto-orbital
effect for 
the relatively narrow p-GaAs quantum well systems (width $\sim$ 150 \AA)
used in ref. [7] is the enhanced conductivity mass at higher magnetic
field values --- the condition $l \ll$ 150 \AA $\;$  necessary for strong
magneto-orbital correction is satisfied for $B \gg$ 4 T whereas the
spin polarization saturation field $B_s$ for the low hole densities
used in ref. [7] is $B_s \sim 3-6$ T; thus the magneto-spin effects
dominate for $B$ upto $3-6$ T whereas the magneto-orbital effects
dominate at higher fields; (5) the magneto-orbital effects are
``similar'' in both cases since both $\rho$ and $\rho_D$ are
proportional to the field-dependent effective mass (which increases
quadratically with the applied field). We mention that these five
features are in excellent qualitative agreement with experimental
results [7].

\begin{figure}
\epsfysize=1.8in
\centerline{\epsffile{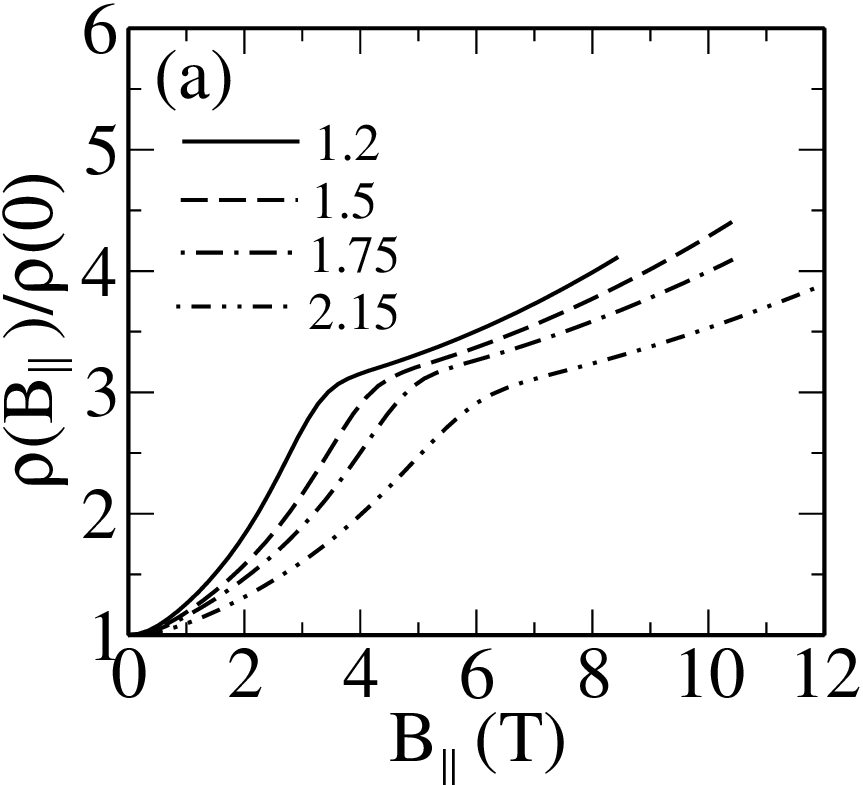}}
\centerline{\epsffile{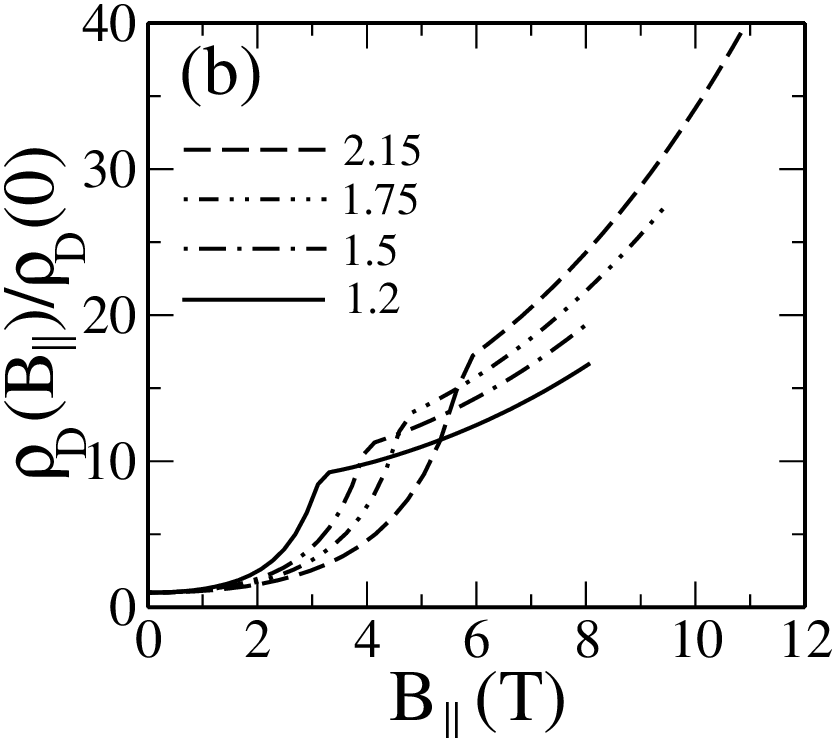}}
\caption{Calculated (a) magnetoresistance $\rho(B_{\|})$ and (b)
  magnetodrag $\rho_D(B_{\|})$ at a temperature $T=100$ mK for
  different density $p=$1.2, 1.5,1.75, 2.15$\times 10^{10}
  cm^{-2}$ as a function of parallel field
  $B_{\|}$.  
\label{Fig2}
}
\end{figure}

In Figs. 1 -- 4 we show our calculated results for $\rho$ and $\rho_D$
within the RPA-Boltzmann Fermi liquid theory. Our theory incorporates
all realistic effects [9-11] with the charged impurity density ($n_i$)
determining $\rho$ as the only unknown free parameter. Our results in
Figs. 1 and 2, where $\rho(B)$ and $\rho_D(B)$ are shown for different
temperatures and different densities, respectively, bear excellent
qualitative resemblance to the corresponding experimental results in
ref. [7]. We are not claiming quantitative agreement with experiment
by any means since our theory is necessarily approximate at the low
carrier densities used in ref. [7] since no exact description of
correlation effects at low densities exists for interacting quantum
Coulomb system of interest here. The qualitative agreement between
theory and experiment is, however, obviously apparent even on a casual
comparison between our Figs. 1 and 2 and the corresponding Figs. 1 and
2 in ref. [7]. In particular, both theory and experiment manifest
qualitatively similar, but by no means identical, behaviors in
$\rho(B)$ and $\rho_D(B)$, arising, as argued above, from the
magneto-spin and magneto-orbital effects.

It is worthwhile to theoretically consider the orbital and the spin
effects separately. (Experimentally this cannot, of course, be done
but one could get some approximate idea about the relative behavior of
the magneto-spin and the magneto-orbital effects in $\rho(B)$ and
$\rho_D(B)$ by concentrating on the `low' $B$ ($<B_s$) and the `high'
$B$ ($>B_s$) regimes, respectively.) In Fig. 3  
we show the
calculated $\rho(B)$ and $\rho_D(B)$ including only the magneto-spin
or only the magneto-orbital effects. Again the importance of the
`low-field' magneto-spin and the `high-field' magneto-orbital effects
on both $\rho(B)$ and $\rho_D(B)$ are manifestly obvious in our
theoretical results.

\begin{figure}
\epsfysize=1.8in
\centerline{\epsffile{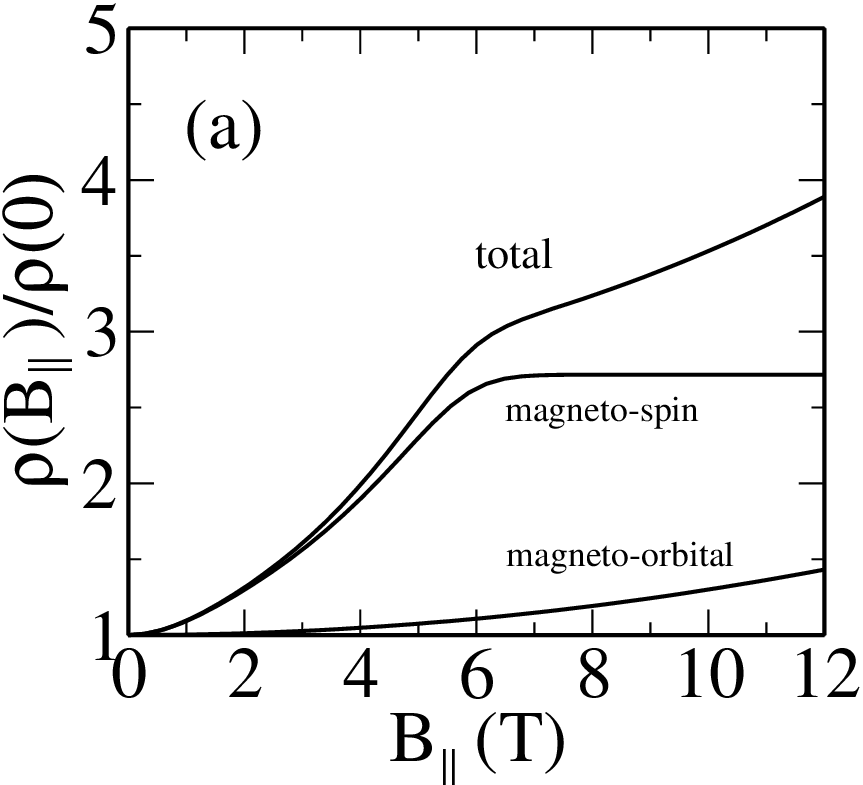}}
\centerline{\epsffile{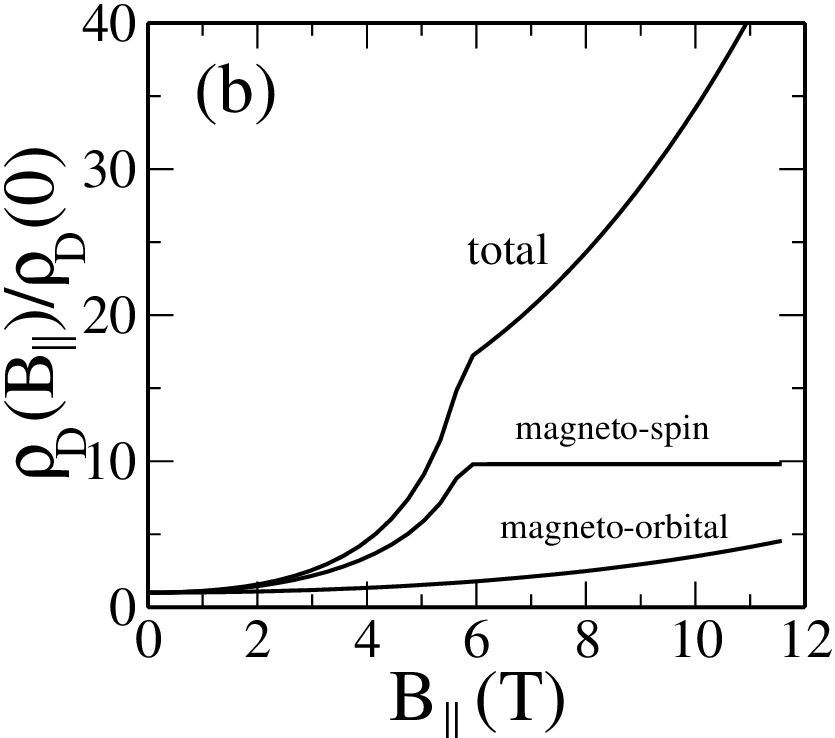}}
\caption{
Calculated (a) $\rho(B)$ and (b) $\rho_D(B)$ including only the magneto-spin
or only the magneto-orbital effects for a hole density $p=2.15\times
10^{10} cm^{-2}$ at $T=100mK$. 
\label{Fig3}
}
\end{figure}

Finally in Fig. 4 we present some clear-cut theoretical predictions
for the temperature 
dependence of bilayer magnetodrag
$\rho_D(B;n,T)$ in the presence of the in-plane magnetic field $B$. In
particular, we fit the temperature-dependence of $\rho_D$ at a fixed
low density 
to approximate power law behaviors: $\rho_D(T) \sim
T^{\alpha}$
with the magnetic field
dependent exponents $\alpha(B)$ 
indicating the nature of the temperature dependence 
of magnetodrag at various magnetic field values. The striking theoretical
prediction, which stands out in Fig. 4, is that $\alpha$ 
manifests a very strong magnetic field dependence with $\alpha(B)$
decreasing from a low-field value of about 2.3 (for $B < B_s$) to a
high-field value of about 1.8 (for $B > B_s$).
This sharp drop in the temperature 
exponent of magneto-drag is a direct
consequence of the strong suppression in magneto-screening arising
from the carrier spin polarization induced by the in-plane magnetic
field. (Note that these exponents are `effective' exponents and not
exact exponents.) Our approximate analysis of the experimental data
[7] indicate that our theoretical results for $\alpha$ 
shown in Fig. 4 are in excellent qualitative (and reasonable
quantitative) agreement with ref. [7], where $\alpha$ changes from
around 2.5 for $B\sim 0$ to about 1.3 for large $B$ 

\begin{figure}
\epsfysize=1.6in
\centerline{\epsffile{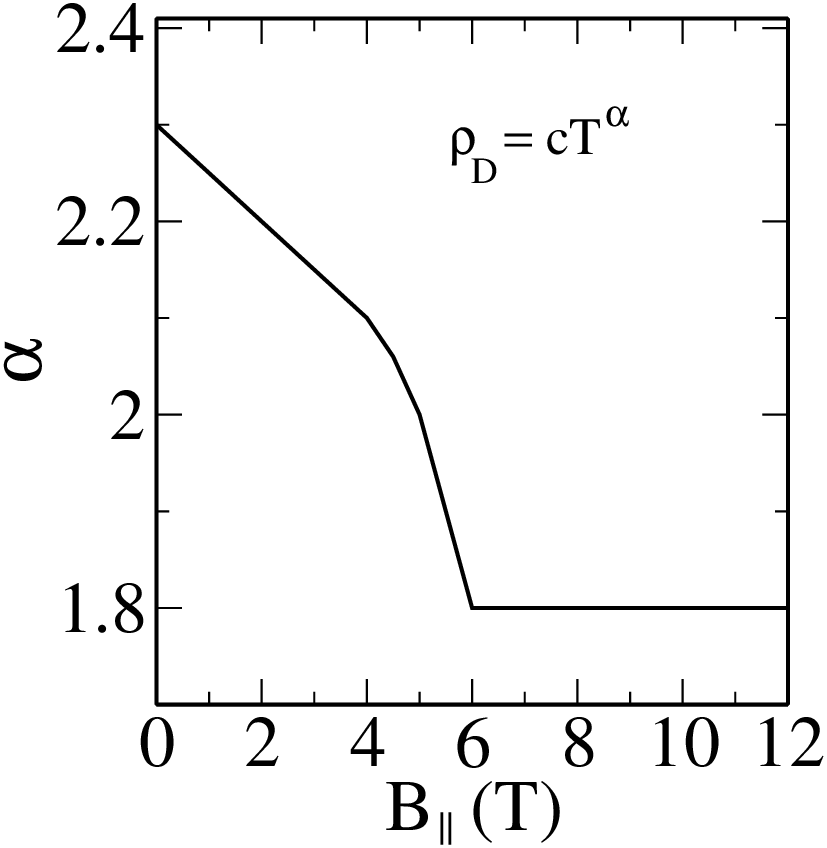}}
\caption{$\alpha$ vs. $B_{\|}$ for hole density $p=2.15 \times 10^{10}
  cm^{-2}$, where the exponent $\alpha(B_{\|})$ is
  deduced from the linear fit of calculated magnetodrag,
  $\rho_D(B_{\|},T) = c T^{\alpha}$. 
\label{Fig4}
}
\end{figure}

We conclude by emphasizing that our Fermi liquid theory based detailed
calculations are in excellent qualitative agreement with the
experimentally observed magneto-drag data [7], and therefore more
exotic non-Fermi liquid theory [15] descriptions (which cannot
typically produce quantitative results as shown in our Figs. 1---4) seem
unnecessary. The `smoking gun' breakdown of the Fermi liquid
description of bilayer drag experiment would be the observation of a
drag resistance which remains finite as $T \rightarrow 0$ since within
the Fermi liquid theory $\rho_D(T=0)=0$. All existing experimental
data seem to be consistent with the Fermi liquid conclusion that
$\rho_D(T\rightarrow 0) \rightarrow 0$.

This work is supported by the US-ONR, the NSF-ECS, the ARO, the ARDA,
and the LPS.

\end{document}